\documentclass[aps,amssymb,amsmath,amsbsy,amsfonts,twocolumn,pra,showpacs]{revtex4}

\usepackage{graphicx}
\usepackage{dcolumn}
\usepackage{bm}
\usepackage{bbm}
\usepackage{psfrag}
\usepackage{epsfig}
\usepackage{amsthm}
\usepackage{color}
\usepackage{mathbbol}
\usepackage{subfigure}
\usepackage{epstopdf,txfonts,hyperref}

\usepackage{bbold}
\usepackage{latexsym}

\usepackage{hyperref}
\usepackage{graphicx}
\usepackage{graphics}
\usepackage{color}

\usepackage{multirow}

\newtheorem{proposition}{Proposition}

\renewcommand{\c}[1]{\mathcal{#1}}

\newcommand{\1}{\mathbbm{1}} 
\newcommand{\idty}{\1}

\DeclareMathOperator*{\tr}{Tr}

\newcommand{\<}{\langle}
\renewcommand{\>}{\rangle}

\newcommand{\beq}{\begin{equation}}
\newcommand{\eeq}{\end{equation}}
\newcommand{\ket}[1]{\ensuremath{\left|{#1}\right\rangle}}
\newcommand{\bra}[1]{\ensuremath{\left\langle{#1}\right |}}
\renewcommand{\rho}{\varrho}

\begin{document}

\title{Discord of response}
\author{W. Roga$^1$, S. M. Giampaolo$^{2}$, and F. Illuminati$^{1,3,4}$\footnote{Corresponding author: illuminati@sa.infn.it}}
\affiliation{$^1$ Dipartimento di Ingegneria Industriale, Universit\`a degli Studi di Salerno,
Via Giovanni Paolo II 132, I-84084 Fisciano (SA), Italy}
\affiliation{$^2$ University of Vienna, Faculty of Physics, Boltzmanngasse 5, 1090 Vienna, Austria}
\affiliation{$^3$ CNISM Unit\`a di Salerno, I-84084 Fisciano (SA), Italy}
\affiliation{$^4$ INFN, Sezione di Napoli, Gruppo collegato di Salerno, I-84084 Fisciano (SA), Italy}

%
%

\date{July 14, 2014}

\begin{abstract}
The presence of quantum correlations in a quantum state is related to the state's response to local unitary perturbations. Such response is quantified by the distance between the unperturbed and perturbed states, minimized with respect to suitably identified sets of local unitary operations. In order to be a {\em bona fide} measure of quantum correlations, the distance function must be chosen among those that are contractive under completely positive and trace preserving maps. The most relevant instances of such physically well-behaved metrics include the trace, the Bures, and the Hellinger distance. To each of these metrics one can associate the corresponding \emph{discord of response}, namely the trace, or Hellinger, or Bures minimum distance from the set of unitarily perturbed states. All these three discords of response satisfy the basic axioms for a proper measure of quantum correlations. In the present work we focus in particular on the Bures distance, which enjoys the unique property of being both Riemannian and contractive under completely positive and trace preserving maps, and admits important operational interpretations in terms of state distinguishability. We compute analytically the Bures discord of response for two-qubit states with maximally mixed marginals and we compare it with the corresponding Bures geometric discord, namely the geometric measure of quantum correlations defined as the Bures distance from the set of classical-quantum states. Finally, we investigate and identify the maximally quantum correlated two-qubit states according to the Bures discord of response. These states exhibit a remarkable nonlinear dependence on the global state purity.
\end{abstract}

\pacs{03.67.Mn, 03.65.Ud, 03.65.Ta}

\maketitle

\section{Introduction}

The characterization, quantification, and physical interpretation of quantum correlations more general than entanglement have been the subject of intensive investigation in recent years~\cite{Modi2012}.
Early seminal works have identified quantum correlations between two parties, $A$ and $B$, in a bipartite quantum state by introducing the \emph{quantum discord}, namely the difference between the two classically equivalent, but quantum inequivalent, entropic formulas for the mutual information~\cite{Modi2012,Zurek2000,Ollivier2001,Henderson2001}. It occurs that the discord is in general non vanishing not only for entangled states, but also for any separable state $\rho_{AB}$ which is not locally convertible via a local change of basis to a so-called \emph{classical-quantum state}, namely a state of the form
\begin{equation}
\rho_{AB}^{(cq)}=\sum_i\ket{i_A}\bra{i_A}\otimes\rho_{B}^{(i)} \, ,
\label{classicalquantum}
\end{equation}
where $\{ \ket{i_A} \}$ is the set of states forming an orthonormal basis in the state space of subsystem $A$ and the $\rho_{B}^{(i)}$ are arbitrary states in the state space of subsystem $B$.

On general grounds, any \emph{bona fide} measure of quantum correlations must satisfy the following minimal set of axioms~\cite{Modi2012,Ciccarello2013,BlindMetrology2013}:

\begin{itemize}
\item I) it must vanish if and only if  the state is a classical-quantum state of the form (\ref{classicalquantum})  (faithfulness criterion);
\item II) it must be invariant under local unitary transformations;
\item III) it must be non-increasing under local, completely positive and trace preserving (CPTP) maps (quantum channels) acting on subsystem $B$;
\item IV) it must reduce to an entanglement monotone if $\rho_{AB}$ is a pure state.
\end{itemize}

The entropic quantum discord satisfies all these axioms. On the other hand, its evaluation requires the highly nontrivial minimization of the entropic functions over all possible local Positive Operator-Valued Measure (POVM) measurements on part $A$, which implies challenging computational difficulties. This drawback has motivated the search for alternative measures of quantum correlations satisfying the basic axioms and at the same time being computationally tractable~\cite{Ollivier2001,Modi2010,Daki'c2010,Streltsov2013,Aaronson2013}. A further, important aspect in the study of quantum correlations concerns their operational interpretation, namely the role that they might play as quantum resources with no classical counterpart in protocols of quantum technologies, ranging from quantum computation and information to metrology and sensing~\cite{Horodecki2003,Datta2008,Luo2008,Madhok2011,Cavalcanti2011,Streltsov2011,Ciccarello2012,Gharibian2012,Girolami2013}. The operational meaning and the reliability with respect to the basic axioms are thus the relevant guiding principles that need to be considered when looking for computable measures of quantum correlations.

The fact that the entropic discord vanishes on the classical-quantum states has been used to define a \emph{geometric} version of the discord, $\delta_G(A|B)$, which characterizes the nonclassical features of a state $\rho_{AB}$ by quantifying its Hilbert-Schmidt distance from the set of classical-quantum states, in complete analogy with the geometric measures of entanglement defined in terms of distances from the set of separable states~\cite{Daki'c2010,Girolami2012,Nakano2012}. Operational interpretations of the geometric discord have been proposed in terms of remote quantum state preparation and the entanglement that is created between a given quantum system and an apparatus performing a measurement on it~\cite{Piani2011,PianiAdesso2012,Streltsov2012,Nakano2012}. Indeed, if $\rho_{AB}$ is a classical-quantum state, and only in this case, there always exists a local orthogonal projective measurement that leaves the state unchanged. This means that the states with non-zero discord are necessarily changed by a local projective measurement. The same observation applies if one replaces local measurements with local unitary operations: \emph{bona fide} measures of quantum correlations can be defined by considering the set of all local unitary transformations $U_A$ that cannot reduce to the identity, namely those local unitaries with fixed, fully non-degenerate spectrum of eigenvalues as, for example, the spectrum of distinct roots of the unity. The local unitaries falling in this class necessarily perturb quantum states with nonvanishing discord~\cite{Streltsov2013,Gharibian2012}.

A welcome feature of the method based on the response of a quantum state to local unitary perturbations is that it allows to introduce a unified approach to the quantification of entanglement and quantum correlations. Indeed, the bipartite \emph{entanglement of response} of pure states of composite quantum systems can be quantified by the change induced by least perturbing local unitary operations and can then be extended to a faithful, full entanglement monotone for mixed states by the convex roof construction~\cite{Monras2011} (see also~\cite{Giampaolo2007,Giampaolo2008,Giampaolo2009,Gharibian2009} for earlier related work).

In the present paper we generalize the entanglement of response to the discord of response for (bipartite) mixed states:
\begin{equation}
D_{R}^{x}(\rho _{AB})\equiv \min_{U_{A}}{\cal{N}}_x^{-1}d_{x}^{2}\left( \rho_{AB},\widetilde{\rho }_{AB} \right) \, ,
\label{DiscOfResp}
\end{equation}
where the index $x$ denotes the possible different types of well behaved, contractive metrics under completely positive and trace-preserving
(CPTP) maps. The normalization factor ${\cal{N}}_x$ depends on the given metrics and is chosen in such a way to assure that $D_{R}^{x}$
varies in the interval $[0,1]$. Finally, $\widetilde{\rho }_{AB} = U_{A} \rho _{AB} U_{A}^{\dagger }$ denotes the unitarily perturbed state,
and the set of local unitary operations $\{U_A\}$ includes all and only those local unitaries with fully nondegenerate spectrum of the complex roots of the unity: $\{e^{i2k\pi/d_A}\}_{k=0}^{d_A-1}$, where $d_A$ is the Hilbert-space dimension of the unitarily perturbed subsystem. Since it is fully non-degenerate and with equal spacing between eigenvalues, we will denote this spectrum as the {\em harmonic spectrum}. The harmonic spectrum is a particular case of fully nondegenerate spectra of complex unimodular numbers.

The most important and physically relevant distances $d_x$ contractive under CPTP maps include the trace, the Hellinger, and the Bures distance.

The trace distance $d_{Tr}$
between any two quantum states $\rho_1$ and $\rho_2$ is defined as:
\begin{equation}
d_{Tr}\left( \rho_1, \rho_2 \right) \equiv \tr \left[ \sqrt{ \left( \rho_1 - \rho_2 \right)^{2}} \right] \, .
\label{tracedistance}
\end{equation}
The Hellinger distance is defined as:
\begin{equation}
d_{Hell}\left( \rho_1, \rho_2 \right) \equiv \sqrt{ \tr \left[ \big( \sqrt{\rho_1} - \sqrt{\rho_2} \big)^{2} \right] } \; \, .
\label{helldistance}
\end{equation}
Finally, the Bures distance, directly related to the Uhlmann fidelity ${\cal{F}}$, is defined as:
\begin{equation}
d_{Bu}\left( \rho_1, \rho_2 \right) \equiv \sqrt{ 2 \left( 1 - \sqrt{ {\cal{F}}(\rho_1,\rho_2) } \right) } \; \, ,
\label{Buresdistance}
\end{equation}
where the Uhlmann fidelity ${\cal{F(\rho_1,\rho_2)}} \equiv\left(\tr\sqrt{\sqrt{\rho_1}\rho_2\sqrt{\rho_1}}\right)^2$.
For each discord of response, trace, Hellinger, and Bures, the normalization factor in Eq.~(\ref{DiscOfResp}) is, respectively:
${\cal{N}}_{Tr}^{-1} = 1/4$, ${\cal{N}}_{Hell}^{-1} = {\cal{N}}_{Bu}^{-1} = 1/2$.

The trace, Hellinger, and Bures discords of response provide different quantifications of the response of a quantum state to least-disturbing local unitary perturbations and satisfy all the previously listed four basic axioms that must be obeyed by a \emph{bona fide} measure of quantum correlations: they vanish if and only if $\rho _{AB}$ is a classical-quantum state; they are invariant under local unitary operations; they are contractive under CPTP maps on subsystem $B$, i.e. the subsystem that is not perturbed by the local unitary operation $U_A$; and they reduce to an entanglement monotone for pure states, for one of which they also assume their maximum possible value ($1$).

In this paper we will focus in detail on the study of the Bures discord of response:
\begin{equation}
D_{R}^{Bu}(\rho_{AB})\equiv\min_{U_A}\frac{1}{2} d_{Bu}^2(\rho_{AB},U_A\rho_{AB}U_A^{\dagger}) \, ,
\label{def:quantumn}
\end{equation}
where, in the following, we will drop the superscript and we will identify the Bures discord of response with the discord of response {\em tout court}:
$D_{R}^{Bu} \equiv D_{R}$.

Specifically, we will show that the Bures discord of response $D_R$: 1) is a faithful measure of quantum correlations satisfying all the basic axioms; 2) reduces to the entanglement of response for pure states; 3) is endowed with a precise operational meaning in terms of state discrimination; 4) possesses a precise quantitative functional relation with the Bures geometric discord, i.e. the geometric measure of quantum correlations defined as the distance of a quantum state from the set of classical-quantum states; and 5) allows to define non-trivial instances of maximally quantum correlated states with nonlinear dependence on the global state purity. Similar detailed investigations can be carried out also for the other well-defined discords of response (trace and Hellinger) and geometric discords (trace and Hellinger), and a comprehensive study on the characterization, quantification, and comparison of the different discords of response and different geometric discords is due to appear soon~\cite{RSI2014}.

In fact, the Bures metric, besides being contractive, shares the unique property of being locally Riemannian~\cite{Bengtsson} and therefore emerges as a natural tool in characterizing the distinguishability of quantum states and operations. Seen as a distance between neighboring operators, it is immediately related to the Fisher Information and the Cramer-Rao bound in the assessment of parameter estimation and quantum metrology~\cite{Braunstein1994}. These are very important features as, in general, other distances, induced by the Schatten $p$-norms or their $L_p$ infinite-dimensional analogs, are not simultaneously Riemannian and contractive ~\cite{Bengtsson,Perez-Garcia}. For instance, the Hilbert-Schmidt distance is Riemannian but not contractive. As already mentioned, the notable contractive exception is the trace distance ($p=1$) which, however, is not Riemannian. The other important instance of contractive distance already mentioned is the Hellinger distance, which, given any pair of quantum states $(\rho_1,\rho_2)$, is defined as the Hilbert-Schmidt distance between their square roots $\sqrt{\rho_1}$ and $\sqrt{\rho_2}$.
On the other hand, simple functions of the Bures distance and of the Uhlmann fidelity provide exact, \emph{a priori} lower and upper bounds to the trace distance $d_{Tr}$. Being the latter related to the error probability in distinguishing two statistical distributions occurring with equal probabilities, in Sec.~\ref{operational} we will exploit these relations in concrete operational applications of protocols for quantum technologies.


Indeed, the Bures discord of response expresses how well a state perturbed by a local unitary operation can be distinguished from the unperturbed state or, in other words, how well the given local unitary transformation can be detected in a given quantum state.
Only a state possessing some amount of quantum correlations can detect unambiguously any local unitary operation {\em with a completely non-degenerate spectrum.}

This relation between Bures metric and local unitary perturbations has immediate bearing on such protocols of quantum technology as the interferometric power in quantum metrology~\cite{BlindMetrology2013}, quantum reading capacity~\cite{Pirandola2011}, and quantum-enhanced refrigeration~\cite{Correa2013}. As a prerequisite for these and other possible applications in protocols of quantum technology, in the following we will provide the exact analytical expressions of the Bures discord of response for two-qubit states with maximally mixed marginals, we will determine the structure of the maximally quantum correlated two-qubit states at fixed global purity, and we will discuss the relation between the discord of response and other geometric measures of quantum correlations.

The paper is organized as follows. In Sec.~\ref{intro_correl} we introduce the Bures discord of response and discuss its general properties and the role played by the spectrum of the associated local unitary transformations in its definition. Its analytical formula for two-qubit states diagonal in the Bell basis
is given in Sec.~\ref{bellstates} and its relation to the Bures geometric discord is discussed. In Sec.~\ref{analysis} we identify the maximally quantum correlated states with respect to the discord of response as a function of the global state purity. Possible applications of the discord of response in protocols of quantum technology are then discussed in Sec.~\ref{operational}.

\section{Local unitary operations, Bures metric, and quantum correlations}\label{intro_correl}

\subsection{General aspects}

Let us consider a quantum state $\rho_{AB}$ on $\c H_A\otimes\c H_B$ and the set of local unitary transformations $U_A\equiv U_A\otimes \idty_B$ where
$\idty_B$ is the identity operator on $B$ while $U_A$ belongs to the set of unitary operators,
irreducible to the identity, with completely non-degenerate spectrum of complex unimodular numbers  $\{\lambda_1,...,\lambda_{d_A}\}$, where $d_A$ is the dimensionality of subsystem $A$~\cite{Streltsov2013,Gharibian2012}. 

Let us next denote by $\Lambda$ the set of unitary matrices with spectrum $\{\lambda_i\}$.
If the state is pure and is fully separable, i.e. it can be written as a tensor product of a state defined on $A$ and a state defined on $B$, there {\it always} exists at least one local unitary operator $U_A$ that leaves the state $\rho_{AB}$ invariant~\cite{Giampaolo2007,Monras2011}. This fact allows to define the {\em Entanglement of response} $E_{R}$:
\begin{equation}
E_{R}^{\Lambda}(\ket{\phi_{AB}})\equiv1- \max_{U_A\in\Lambda}F(\ket{\phi_{AB}},U_A\ket{\phi_{AB}}) \, ,
\label{stellarentanglement}
\end{equation}
where $F$ is the Uhlmann fidelity function 
which for pure states reduces to (scalar product) overlap. The entanglement of response can then be promoted to a full entanglement monotone for general mixed states via the convex roof extension of Eq.~(\ref{stellarentanglement})~\cite{Monras2011}.

The generalization of this procedure to mixed states leads to the definition of a faithful measure of quantum correlations, the discord of response. The basic ingredient paving the way to this generalization is provided by the following proposition.

\begin{proposition}
\label{propuno}
If and only if $\rho_{AB}$ is a classical-quantum state $\rho_{AB}^{(cq)}$ of the form (\ref{classicalquantum}), there exists at least one local unitary operator $U_{A}\in\Lambda$, in the set of unitary operators with completely non-degenerate spectrum, such that
\begin{equation}
\tilde{\rho}_{AB}\equiv U_{A}\rho_{AB}U_{A}^{\dagger} = \rho_{AB} \, .
\label{eq:main}
\end{equation}
\end{proposition}

The proof is given in the Appendix. From Proposition~\ref{propuno} it follows that the minimum distance of
$\rho_{AB}$ from the set of states $\tilde{\rho}_{AB}= U_A\rho_{AB}U_A^{\dagger}$, where $U_A\in \Lambda$,
satisfies axiom I): it vanishes if and only if $\rho^{AB}$ is classical-quantum. Verification of the remaining axioms II) through IV) depends on the choice of the distance. Since the Bures distance is unitarily invariant and contractive under CPTP maps (either local or global), the following quantity is a proper measure of quantum correlations:
\begin{equation}
D_{R}^{\Lambda}\left(\rho_{AB}\right)=
1-\max_{U_A\in\Lambda}\tr\sqrt{\sqrt{\rho_{AB}}U_A\rho_{AB}U_A^{\dagger}\sqrt{\rho_{AB}}} \, ,
\label{quantumness1}
\end{equation}
where the perturbed subsystem is the one with smaller Hilbert-space dimension, i.e. $d_A \leq d_B$.
This expression defines the (Bures) discord of response. The maximization is performed over the set of all local unitary operations with fully nondegenerate spectrum of complex unimodular numbers, and the normalization is chosen such that the distance does not exceed unity, with the absolute upper bound achieved by the maximally entangled pure states. Moreover, for pure states $D_{R}^{\Lambda}$ is a simple monotonic function of the entanglement of response:
\begin{equation}
D_{R}^{\Lambda}(\ket{\phi})=1-\sqrt{1-E_{R}^{\Lambda}} \; .
\end{equation}
Therefore, the discord of response satisfies all axioms for a proper measure of quantum correlations.

The discord of response $D_{R}^{\Lambda}$, quantifies the distinguishability between quantum states before and after the application of a local unitary operation with nondegenerate spectrum. Equivalently, it quantifies the minimum possible \emph{response} of a composite quantum system in a given state, subject to local unitary perturbations, by identifying the local operation that induces the least perturbing effect. It is important to understand that this type of distinguishability differs in general from the distinguishability usually considered in the study of quantum correlations, namely how well a local measurement can distinguish between elements of a statistical mixture of quantum states of a composite quantum system.

The asymmetry in treating the subsystems in the definition Eq. (\ref{quantumness1}) is an intrinsic feature
of this type of discord, like many other measures of quantum correlations. If we take the analogous definition but with minimization defined under local unitary transformations on both subsystems, then one finds that the resulting measure can vanish for some states that do not belong to the set of classical-quantum states. The simplest relevant example is the one of two-qubit Werner states which are invariant under all the transformations of the form $U_{A}\otimes U_{B}\rho_{AB}^{w}U_{A}^{\dagger}\otimes U_{B}^{\dagger}$, with $U_{A} = U_{B} = U$. It then follows that $U \otimes U \rho_{AB}^{w}U^{\dagger} \otimes U^{\dagger} = \rho_{AB}^{w}$. The would-be symmetric Bures discord is thus always vanishing for these states and does not define a faithful measure of quantum correlations.

\subsection{Choice of the spectrum}

We will now investigate in more detail the role played by the choice of the spectrum of the local unitary operations $U_A\in\Lambda$ that enter in the definition of the discord of response. Extremizing over the entire set of local operations $U_A$ without any restriction on their spectrum implies trivially that the minimum is always given by a vanishing discord of response, Eq. (\ref{quantumness1}), achieved  by the local identity $U_A=\idty$. In order to define a nontrivial discord of response one needs to impose constraints which separate unambiguously the selected classes of unitary transformations from the identity.

Proposition \ref{propuno} guarantees that every choice of a completely nondegenerate spectrum of $\Lambda$ provides a measure of quantum correlations that vanishes if and only if the measured state is classical-quantum. However, each different choice of the spectrum provides a different measure, which can imply different ordering among the states, a situation similar to the case of other measures of nonclassicality, defined in terms of local unitary operations, such as the entanglement of response~\cite{Monras2011} and the local quantum uncertainty~\cite{Girolami2013}.

If the perturbed subsystem is a qubit, $d_A=2$, or a qutrit, $d_A=3$, the condition that $U_A$ is traceless
implies a particular case of fully non-degenerate spectrum of unimodular complex numbers, i.e. the so-called harmonic spectrum of equally spaced complex roots of the unity. The traceless condition provides in these cases the proper separation of $U_A$ from the identity. This is not the case any more when $d_A\geq 4$. In this case the traceless condition does not guarantee that the spectrum is completely non-degenerate, while harmonicity remains as a sufficient condition to this end, and other choices are possible.

On the other hand, the choice of the harmonic spectrum, i.e. with homogeneously spaced eigenvalues, can be strongly motivated and justified on physical grounds. Indeed, the shape of the spectrum becomes an especially delicate problem for higher-dimensional systems ($d_A>>1$). Although any choice of a generic non-degenerate spectrum yields a vanishing discord if and only if the state is classical-quantum, the density and spacing of eigenvalues will determine the main character of the different measures. Suppose we choose an inhomogeneous spectrum. In this case, if there are subspaces associated to very similar eigenvalues of $U_A$, in such subspaces the local unitary transformations will act almost like the identity. The quantum correlations between these subspaces and the rest of the system will then be weighted very poorly by the corresponding discord of response, and will be almost completely hidden. To illustrate this point in more detail, let us consider a composite system $ABCD$, where the state of party $AD$ is maximally entangled,
while the state of party $BC$ is a product state, and we choose a non-degenerate but inhomogeneous spectrum of local unitary operations $U_{AB}$. The minimization over $U_{AB}$ will then tend to minimize the impact of the maximally entangled part acting on $A$. This impact will become ever vanishingly smaller the closer are the eigenvalues of $U_{AB}$, while intuitively one should expect exactly the opposite behaviour, i.e. that the maximally entangled subsystem should give a significant contribution to the quantum discord between $AB$ and $CD$. This example shows that the unitary transformations of non-degenerate but non-uniformly distributed spectrum introduces non-equivalent weights to the correlations in different subspaces of the system.
To avoid such spurious situations the only sensible choice appears to be that of the harmonic spectrum.

A further operationally motivated argument in favor of the harmonic spectrum will be given in Sec.~\ref{operational}, where we will show its optimality for the estimation of the maximal error probability in quantum reading protocols. In the following we will always consider the discord of response defined by the minimization over local unitary operations $U_A$ with harmonic spectrum. Therefore from now on we will omit the superscript $\Lambda$ labeling the spectrum in the definition of the discord of response $D_R$.

\section{Discord of response for states with maximally mixed marginals}\label{bellstates}

In this section we will compute explicitly the general expression (\ref{quantumness1}) for the discord of response in the case of some relevant classes of two-qubit states.
Let us now focus on the case in which $A$ and $B$ are qubit subsystems and let us consider a state that admits as eigenvectors of its density matrix the set of the Bell states:
\begin{eqnarray}
|\Psi_{\pm}\>&\equiv&\frac{1}{\sqrt{2}}\left(|0_A1_B\>\pm|1_A0_B\>\right)\label{bell1},\\
|\Theta_{\pm}\>&\equiv&\frac{1}{\sqrt{2}}\left(|0_A0_B\>\pm|1_A1_B\>\right)\label{bell2} \, .
\end{eqnarray}
We will parameterize these two-qubit states $\rho_{\gamma}$ by the vector of eigenvalues $\vec{\gamma}=(\gamma_1, \gamma_2, \gamma_3, \gamma_4)$. One has
\begin{equation}
\rho_{\gamma}=\frac{1}{2}\begin{bmatrix}
 \gamma_1 +\gamma_2  & 0 & 0 & \gamma_1 -\gamma_2  \\
 0 & \gamma_3 +\gamma_4  & \gamma_3 -\gamma_4  & 0 \\
 0 & \gamma_3 -\gamma_4  & \gamma_3 +\gamma_4  & 0 \\
 \gamma_1 -\gamma_2  & 0 & 0 & \gamma_1 +\gamma_2
\end{bmatrix} \, .
\label{matrix}
\end{equation}
Many important instances of two-qubit states belong to this class as, for instance, the Werner states, defined in the Bell basis Eqs. (\ref{bell1}) and (\ref{bell2}). In this basis, the Werner states read
\begin{equation}
\!\rho_w \!\!=\!\!f\ket{\Psi_{\!-}}\!\bra{\Psi_{\!-}}
+\frac{1-f}{3}(\ket{\Psi_{\!+}}\!\bra{\Psi_{\!+}}
+\ket{\Theta_{\!-}}\!\bra{\Theta_{\!-}}+\ket{\Theta_{\!+}}\!\bra{\Theta_{\!+}}) \, .
\label{wernerstates}
\end{equation}
For these states one has $\vec{\gamma}=(f,(1-f)/3,(1-f)/3,(1-f)/3)$.
Since the states $\rho_{\gamma}$ are convex combinations of Bell states, their state reductions to either one of the subsystems are maximally mixed. The analytical expression of the standard geometric discord for this class of states (using the Bures metric) has been reported recently~\cite{Spehner2013a,Spehner2013b}.

The following proposition holds for the discord of response of two-qubit states diagonal
in the Bell basis:

\begin{proposition}
\label{prop_belldiag}
The discord of response $D_{R}(\rho_{\gamma})$ of any two-qubit state
diagonal in the Bell basis is
\begin{eqnarray}
D_{R}(\rho_{\gamma})\!\!\!&=&\!\!\!\min_{c.p.}\left[1-2 \left(\sqrt{\gamma_1  \gamma_2}+ \sqrt{\gamma_3  \gamma_4}\right)\right] \, ,
\label{Buformula}
\end{eqnarray}
where minimization is taken over cyclic permutations of the eigenvalues $\gamma_1, \gamma_2$ and $\gamma_3$.
\end{proposition}

\begin{figure}
\includegraphics[width=9cm]{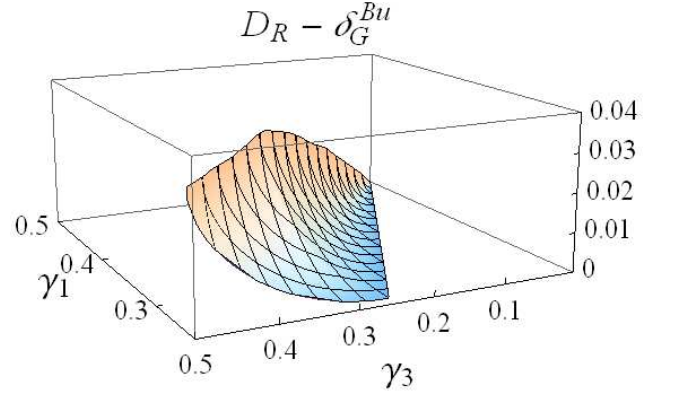}
\caption{Difference between the discord of response $D_{R}$ and the Bures geometric discord $\delta^{Bu}_G$ for the Bell-diagonal states $\rho^{AB}_{\gamma}$ defined in Eq.~(\ref{matrix}), as a function of the eigenvalues $\gamma_1$ and $\gamma_3$. For ease of illustration we have reported the case $\gamma_1=\gamma_2$. In the case of Bell-diagonal states extended numerical analyses show that
the difference is always strictly positive. 
}
\label{bellcomparison}
\end{figure}

The detailed proof is given in the Appendix. Thus equipped, it is possible to compare the Bures discord of response and the Bures geometric discord defined as~\cite{Spehner2013a,Spehner2013b}:
\begin{equation}
\delta_G^{Bu}(\rho_{AB})\equiv \frac{2}{2-\sqrt{2}}\left(1-\max_{\rho_{AB}^{(cq)}}\sqrt{F(\rho_{AB},\rho_{AB}^{(cq)})}\right) \, ,
\end{equation}
where $F$ is the Uhlmann fidelity 
and the maximization is performed over the set of all classical-quantum states, Eq. (\ref{classicalquantum}).
The results of this comparison are illustrated in Fig.~\ref{bellcomparison}, which shows
that the discord of response of Bell-diagonal states always bounds the geometric discord from above.

Werner states, Eq.~(\ref{wernerstates}) are a particularly important instance of states diagonal in the Bell basis. Their eigenvalues are $(f,(1-f)/3,(1-f)/3,(1-f)/3)$, where $f$ is the parameter that selects the type of Werner state in the definition, Eq.~(\ref{wernerstates}). For Werner states, the discord of response is
\begin{equation}
D_{R}\left(\rho_w\right)=1-\frac{2}{3} (1-f)-\frac{2 \sqrt{\left(f-f^2\right)}}{\sqrt{3}} \, .
\end{equation}
We can re-express it as a function of the global state purity $P\equiv \tr{\left(\rho^2\right)}$. For a given purity $P\leq 1/3$, the  parameter $f$ can take two values, which yield two different ranges for
$D_{R}\left(\rho_w\right)$ as a function of $P$:
\begin{eqnarray}
D_{R}\left(\rho_w\right)\Big|_{\mp}&=&1-\frac{1}{6} \left|3\mp\sqrt{12 P-3}\right|\label{plusminus}\\
&-&\frac{\sqrt{\left|-6 P\pm\sqrt{12 P-3}+3\right|}}{\sqrt{6}} \, .
\nonumber
\end{eqnarray}

\begin{figure}
\includegraphics[width=8.0cm]{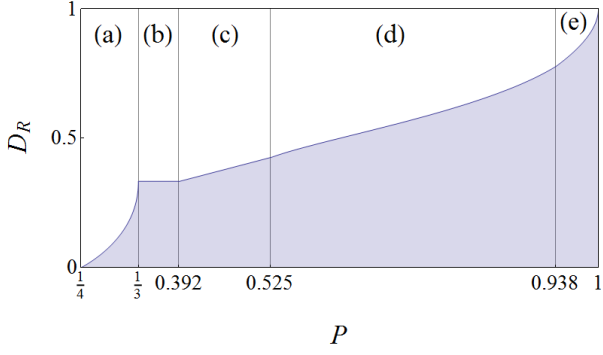}
\caption{Behavior of the upper bounds for the discord of response $D_{R}\left(\rho_{AB}\right)$ of general two-qubit states as a function of the global state purity $P\equiv\tr\left(\rho_{AB}\right)^2$. The shadowed region identifies the set of admissible values of the discord of response and the region's upper boundaries identify the maximally quantum correlated two-qubit states at fixed global state purity. Detailed expressions are discussed in Sec.~\ref{analysis} of the main text. The different types of functions forming the upper boundaries divide the purity range in five distinct regions, as discussed in detail in the main text.}
\label{buresborder}
\end{figure}

The solution $D_{R}\left(\rho_w\right)\Big|_{+}$ holds only for $\frac{1}{4}\leq P\leq \frac{1}{3}$ (i.e. $0\leq f\leq  \frac{1}{4}$),
while  the solution $D_{R}\left(\rho_w\right)\Big|_{-}$ holds for $\frac{1}{4}\leq P\leq1$ (i.e. $\frac{1}{4}\leq f \leq 1$).
Equations (\ref{plusminus}) define the upper bounds, as illustrated in Fig.~\ref{buresborder}, of the admissible values of the discord of response for general two-qubit states as a function of the global state purity in the ranges (a): $\frac{1}{4}\leq P\leq \frac{1}{3}$ and (e): $0.94\leq P\leq 1$.
Therefore, in these ranges Werner states are maximally quantum-correlated. In the remaining ranges of the global state purity maximally quantum correlated two-qubit states belong to classes more general than that of Werner states. This finding is at variance with the result that is obtained using as a measure of quantum correlations the geometric discord based on the Hilbert-Schmidt metric~\cite{Streltsov2012}. Indeed, this fact illustrates the discrepancy between predictions based on the Hilbert-Schmidt and the Bures distance. The former is a linear function of the global state purity; moreover, it is not contractive under CPTP maps and is thus not a \emph{bona fide} measure of quantum correlations. The latter is Riemannian, is a \emph{bona fide} measure of quantum correlations contractive under CPTP maps, and is a nonlinear function of the global state purity.

\section{Maximally quantum correlated states}\label{analysis}

The results of this Section were obtained by extensive numerical analyses performed as follows.
Two-qubit states were generated by choosing randomly their eigenvalues and eigenvectors.
We have generated $10^7$  random  two-qubit states with eigenvectors taken as
columns of a unitary matrix chosen randomly from the set of uniformly distributed unitary matrices.
Candidate upper-boundary states were selected by repeated evaluation of discord and global state purity
following infinitesimal perturbations of the states falling in the vicinity of the upper boundaries.
The emerging structure of maximally quantum-correlated mixed two-qubit states in the entire range of global state purity is characterized as follows: in the ranges (a) and (e) of Fig.~\ref{buresborder}, i.e. $\frac{1}{4}\leq P\leq \frac{1}{3}$ and $0.94\leq P\leq 1$, the maximally quantum-correlated states are Werner states. In the range (b) of Fig.~\ref{buresborder}, i.e. $\frac{1}{3}\leq P\leq 0.39$, the maximally quantum-correlated states are of the following form in the computational basis $\{\ket{i_A,j_B}\}_{i_A,j_B=1}^2$:
\begin{equation}
\rho^{(mq)}_{b}=\frac{1}{6}\begin{bmatrix}
 1  & 0 & 0 &1  \\
 0 & 4-b  & 0  & 0 \\
 0 & 0  & b  & 0 \\
 1  & 0 & 0 & 1
\end{bmatrix} \, ,
\label{matrixb}
\end{equation}
where $b$ is the following nonlinear function of the global state purity: $b=2+\sqrt{6}\sqrt{3P-1}$. Here we pause to observe that the evaluation of the general expression of the discord of response $D_{R}$, Eq.~(\ref{quantumness1}), can be simplified whenever party $A$ is a qubit and $B$ is arbitrary finite-dimensional system. The square root of the fidelity between $\rho_{AB}$ and $U_A\rho_{AB}U_A^{\dagger}$ is $\sqrt{F}=\tr{\sqrt{\sqrt{\rho_{AB}}U_A\rho_{AB}U_A^{\dagger}\sqrt{\rho_{AB}}}}
=\tr{\sqrt{\sqrt{\rho_{AB}}U_A\sqrt{\rho_{AB}}\sqrt{\rho_{AB}}U_A^{\dagger}\sqrt{\rho_{AB}}}}$,
equivalent to the sum of the absolute values of the eigenvalues of  $\sqrt{\rho_{AB}}U_A\sqrt{\rho_{AB}}$ which, by having the same eigenvalues, is a matrix similar to $\rho_{AB}U_A$. Therefore, the discord of response $D_{R}$ takes the general form
\begin{equation}
D_{R}\left(\rho_{AB}\right)=1-\max_{U_A\in\Lambda}
\sum_i\left|\xi_i\left(\rho_{AB}U_A\right)\right| \, ,
\label{simplification}
\end{equation}
where $\xi_i\left(\rho_{AB}U_A\right)$ are the eigenvalues of the matrix $\rho_{AB}U_A$.
Using Eq.~(\ref{simplification}) it is straightforward to conclude that
the discord of response $D_{R}(\rho^{(mq)}_b)$ is constant for all values of the global state purity
in the range $ \frac{1}{3} \leq P \leq 0.39$ and takes the value
$D_{R}(\rho^{(mq)}_b) = \frac{1}{3}$. In this case we can easily identify the decomposition of $\rho^{(mq)}_b$ onto the mixture of a maximally entanglement state (external block) and a separable state (the internal block)~\cite{Lewenstein1998}, and the discord of response depends only on the entangled term of the decomposition.

In the region (c), corresponding to the range of global state purity $0.39\leq P\leq 0.53$, we can identify the states compatible with the upper boundary of the discord of response up to two numerical parameters $a_1$ and $a_2$. The corresponding maximally quantum-correlated states are the following rank 3 states written in the computational basis $\{\ket{i_A,j_B}\}_{i_A,j_B=1}^2$:
\begin{eqnarray}
\rho^{(mq)}_{c}=
\frac{1}{6}\left[\begin{smallmatrix}
K  & 0 & 0 &\sqrt{K(c-K)}  \\
0 & (1+4K-3c)/2  & 0  & 0 \\
0 & 0  & (1-4K+c)/2  & 0 \\
\sqrt{K(c-K)}  & 0 & 0 & c-K
\end{smallmatrix}\right] \, ,
\label{matrixc}
\end{eqnarray}
where $K=a_1+a_2P$, the parameters $a_1\approx 0.03,\ a_2\approx 0.35$ are numerical constants, and
$c=\frac{1}{7}(1+8K+\sqrt{2}\sqrt{7P-24K^2+8K-3})$. The discord of response of these states is the following nonlinear function of the global state purity:
\begin{eqnarray}
&&\!\!\!\!\!\!\!D_{R}\!\left(\!\rho^{(mq)}_{c}\!\right)\!\approx\!0.3 P\!-\!0.35 \sqrt{(2.8\!-\!P)(P\!-\!0.34)}\nonumber\\
&&+0.88-1.7 \Big[0.013   P^2+0.19\Big.\nonumber\\
&&+\left(0.09 \sqrt{(2.8-P)(P-0.34)}-0.09\right) P\nonumber\\
&&\left.-0.2\sqrt{(2.8-P)(P-0.34)}\right]^{\frac{1}{2}} \, .
\end{eqnarray}

In the region (d) of Fig.~\ref{buresborder}, corresponding to the range of values of the global state purity $0.53\leq P\leq 0.94$, the maximally quantum-correlated states are rank 2 states of the following form in the computational basis:
\begin{equation}
\rho^{(mq)}_{d}=\left[
\begin{matrix}
 (1-d) \cos ^2(\eta) & 0 & 0 &
   (1-d) \cos (\eta) \sin
   (\eta) \\
 0 & 0 & 0 & 0 \\
 0 & 0 & d & 0 \\
 (1-d) \cos (\eta) \sin
   (\eta) & 0 & 0 & (1-d) \sin
   ^2(\eta)
\end{matrix}
\right]
\label{matrixd},
\end{equation}
where $d=\frac{1}{2} \left(1-\sqrt{2 P-1}\right)$ and
{\small
\begin{equation}
\eta=\frac{1}{2} \cos ^{-1}
\left(\frac{2
   P+\sqrt{(1-P) \left(-P+2 \sqrt{2
   P-1}+3\right)}-2}
{-\frac{1}{2}\left(\sqrt{2
   P-1}+1\right)^2}\right).
\end{equation}
}

We can summarize the results of our analysis as follows. According to the
discord of response, the maximally quantum-correlated two-qubit states at fixed global state
purity are:

\begin{eqnarray*}
&&(a)\ for\ \frac{1}{4}\leq P\leq \frac{1}{3}\ Werner\ states\ with\ f\leq \frac{1}{4}, \\
&&(b)\ for\ \frac{1}{3}\leq P\leq 0.39\ states\ of\ the\ form\ Eq.\  (\ref{matrixb}),\\
&&(c)\ for\ 0.39\leq P\leq 0.53\ states\ of\ the\ form\ Eq.\  (\ref{matrixc}),\\
&&(d)\ for\ 0.53\leq P\leq 0.94\ states\ of\ the\ form\ Eq.\  (\ref{matrixd}),\\
&&(e) \ for\ 0.94\leq P\leq 1\ Werner\ states\ with\  f \geq 0.97 \, .
\end{eqnarray*}

In conclusion, maximally quantum correlated states with respect to the discord of response possess a rich structure that is due to the fact that the Bures distance is a nonlinear function of the global state purity. Changing the global purity induces marked discontinuities in the rank and the form of the maximally quantum correlated states. On the other hand, they can be always represented in the same computational basis in the form of block diagonal density matrices. It should be noted that analogous block-diagonal structures and discontinuities in the rank are the same features shared by maximally entangled mixed states~\cite{Munro2001,Wei2003}.

\section{Operational aspects and applications}\label{operational}

The discord of response is quantified by the Bures distance between a given input state and the output state obtained from it under the action of a local unitary perturbations. From the operational interpretation of the Uhlmann fidelity and Bures metric it stems that the more quantum-correlated a quantum state, the more such state is sensitive to a local unitary perturbation, and the better it can be distinguished from the perturbed state. Therefore, it seems possible to control or \emph{tame} an external environment by suitable local unitary quantum driving of statistical noise and thermal fluctuations.

These features can be of immediate interest in the context of protocols of quantum technology where distinguishability with respect to local unitary transformations plays a significant role. The detailed theoretical and experimental study of the application of the discord of response to protocols of quantum technology such as e.g. quantum reading and illumination, quantum metrology, and related problems in quantum sensing and quantum technologies is beyond the scope of the present work, which is of general and mathematical nature, and will presented elsewhere~\cite{inpreparation,inpreparation2}. In the following we will only present a brief discussion of how the discord of response can be fruitfully applied to the study of two such protocols of quantum technology. Specifically, we will consider the problem of
the quantum reading capacity~\cite{Pirandola2011}, in which one needs to distinguish between two or more unitary quantum channels, and a problem of quantum parameter estimation, in which not only the parameter is unknown but also the local driving Hamiltonian is not completely known~\cite{BlindMetrology2013}.

The idea of reading classical data by means of quantum states, yielding a significant advantage over purely classical resources has been recently proposed in Ref.~\cite{Pirandola2011}. The data are written on a classical device (in the optical implementation it is a
CD-like device) by means of different types of cells. The quantum transmitter which has to extract the encoded information is prepared in an initial state. Passing a cell the transmitter changes its properties in a way depending on the state of the cell, and it is detected.
The task is to recognize which cell has occurred based on the output state of the transmitter.
Therefore the problem of reading is reduced to the problem of distinguishing the different output states of the transmitter.

The most common implementations are based on optical technologies, see e.g.~\cite{Hirota2011}.
In this setting, one has to distinguish between two main coding scenarios depending on the type of transmitters and channels that are used. The first scenario is called "amplitude shift keying" (ASK) in which the changes in the state of the transmitter are caused by
the cell-dependent losses of the intensity of the transmitted signal. The second scenario is called "phase shift keying" (PSK). This is a coding without energy losses, which however demands a very high coherence of the transmitter.

If the transmitter is quantum, the cells play the role of quantum channels. The ASK scenario is an instance of dissipative channel coding, while in PSK we have a coding by means of the unitary transformations.
It has been shown, particularly in the low energy regime, that the transmitters which are quantum  (entangled, squeezed etc.) can provide some advantages over the classical states (convex combinations of coherent states) in both the ASK and PSK scenarios~\cite{Pirandola2011,Hirota2011}.

The probability of error in the discrimination between the output states of the transmitter after passing through equiprobable channels, $\Phi_1$ or $\Phi_2$ is provided by the Helstrom formula
$$
P_{err}(\rho,\Phi_1,\Phi_2)=\frac{1}{2}(1-\frac{1}{2}D_{Tr}\big(\Phi_1(\rho),\Phi_2(\rho)\big) \, ,
$$
where $D_{Tr}$ is the trace distance. In the binary, loss-free coding only two channels are needed. Without loss of generality we can assume that one of them is the identity and the second one is a local unitary transformation $U_A$.
For a given transmitter $\rho_{AB}$ we can introduce the device-independent characteristic of the quantum reading by assuming the worst case scenario and maximizing the error probability over all local unitary operations $U_A$ with non-degenerate spectrum.
One then obtains the probability of error expressed in terms of the discord of response defined by means of the trace distance. This type of discord of response characterizes the difference  between the maximal probability of error in the above scenario and the classical head-tail value 1/2, i.e. the absolute maximum of the error probability.

At this point let us discuss again the role played by the choice of the spectrum of the local unitary operations in the definition of the discord of response. The following proposition shows that if subsystem $A$ is a qubit, then the choice of the harmonic spectrum yields the smallest possible maximal probability of error. This in turn implies that the choice of the harmonic spectrum is the optimal one in order to code messages via unitary operations:

\begin{proposition}
\label{proptr2}
{\rm Consider an arbitrary bi-partite transmitter state $\rho^{AB}$, where subsystem $A$ is a qubit, $d_A=2$, and subsystem $B$ is arbitrary finite-dimensional. Assume that we encode the message using the identity channel $\idty_A$ and an arbitrary unitary channel $U_A$, both occurring with equal probabilities. Then, the maximal probability of error of discrimination between the outputs of the operations given by $\idty_A$ and by $U_A$
is smallest if the spectrum of $U_A$ is harmonic.}
\end{proposition}

The proof is given in Appendix~\ref{appprop2}. At the moment there is no proof for $d_A>2$ and the task appears challenging.  The result of the above Proposition suggests that the choice of the harmonic spectrum for the local unitary operations that define the discord of response provides the optimal estimation of some crucial figure of merit in quantum reading protocols. Indeed, since the trace distance $D_{Tr}$ is challenging to evaluate in many relevant instances, for example in the case of repeated sampling~\cite{Audenaert2007} or in the case of Gaussian probe states~\cite{Pirandola2008}, useful estimators such as the quantum Chernoff bound~\cite{Audenaert2007} and the Bures distance $D_{Bu}$ can come at hand. In particular, the latter allows to derive lower and upper bounds on the error probability thanks to the following relations to the trace distance:
\begin{equation}
D_{Bu}(\rho_1,\rho_2)^2\leq D_{Tr}(\rho_1,\rho_2)\leq 2 D_{Bu}(\rho_1,\rho_2) \, .
\label{burestracerelation}
\end{equation}
Indeed, these two distances are topologically equivalent~\cite{Fuchs1999,Belavkin2005}. This means that if two states become closer to each other with respect to one of these measures they are also closer to each other with respect to the other measure. Moreover, the bounds in Ineq.~(\ref{burestracerelation}) are simple functions of the quantum Chernoff bound if one of the states is pure~\cite{Pirandola2008}. Moreover, the
discord of response allows to show the crucial advantage of quantum reading over the classical reading. In particular, our results imply that the states with large discord of response are able to read any coding by means of local unitary operations, while this is impossible for classical-quantum transmitters, i.e. transmitters with vanishing discord of response.

The results of the present paper may also find interesting applications in the context of metrology, particularly for interferometry and phase estimation~\cite{BlindMetrology2013}. In such problems one has to estimate some unknown phase $\phi$ introduced at one arm of an interferometer by a unitary transformation $e^{i\phi H}$, while the generating Hamiltonian $H$ is either unknown or characterized only by its spectrum.
The precision of the phase determination is estimated by the interferometric power, a function of the quantum Fisher information $F$ dependent on the probing state and the generating Hamiltonian $H$. Not knowing explicitly $H$, one has to minimize the Fisher information over all Hamiltonians $H$ in order to characterize the precision of the phase estimation. For arbitrary states $\rho_{AB}=\sum_iq_i\ket{\phi_i}\bra{\phi_i}$ the quantum Fisher information
$F(\rho_{AB},H_A)$ reads
$$
F(\rho_{AB},H_A)=4\sum_{i<j:q_i+q_j\neq 0}\frac{(q_i-q_j)^2}{q_i+q_j}|\<\phi_i|H_A\otimes\idty_B|\phi_j\>|^2 \, ,
$$
and the associated interferometric power reads
$$
\c P^A(\rho_{AB})\equiv \frac{1}{4}\min_{H_A} F(\rho_{AB},H_A) \, .
$$
It has been recently found that $\c P^A(\rho_{AB})$ is nonvanishing only for discordant states and can be suitably maximized, e.g. with experimental setups based on NMR technology~\cite{BlindMetrology2013}. In this scenario, for local generating Hamiltonians $H_A$ with non-degenerate spectrum we can characterize the sensitivity of the probed state on the action of $e^{i\phi H_A}$ 
by the Bures distance between the state perturbed locally by $e^{i\phi H_A}$ and the unperturbed state after minimization over all local Hamiltonians $H_A$ with fixed non-degenerate spectrum. Such a measure vanishes if and only if there exists a local Hamiltonian which does not perturb the probing state. As a consequence, the results presented in Sec.~\ref{analysis} of the present work identify the states of given purity which are most sensitive to the action of unknown local Hamiltonians. These qualitative similarities between the discord of response and the interferometric power need to be supplemented by more extended studies in order to investigate stricter quantitative relations between them.

{\em Note added in proof:} After our work was completed, a related paper by Farace et. al. has appeared~\cite{Farace2014}. In this work the authors introduce, in the context of protocols of quantum illumination, a further response-based measure of non-classical correlations, the so-called {\em discriminating strength}. This measure is quite similar in spirit to the discord of response, being defined through the Bures distance with respect to a local unitary perturbation. The main difference between the discord of response and the discriminating strength lies in the fact that in the former the Bures distance is defined through the Uhlmann fidelity while in the latter it is defined through the quantum Chernoff bound. In the perspective of future work along this line of investigation, it will be interesting to analyze analogies and differences between these two response-based measures of quantum correlations.

{\em Second note added in proof:} Very recent collaboration with the authors and Dr. D. Spehner has allowed to prove that an exact analytical relation always holds between the Bures discord of response and the geometric Bures discord whenever subsystem $A$ is a qubit and subsystem $B$ is an arbitrary $d$-dimensional system. Other general analytic relations between different types of discords of response and geometric discords have also been discovered and will be reported in the same comprehensive study~\cite{RSI2014}.


\begin{acknowledgments}
The authors are grateful to the two anonymous Referees for their very competent and useful comments. The authors acknowledge financial support from the Italian Ministry of Scientific and Technological Research under the PRIN 2010/2011 Research Fund, and from the EU FP7 STREP Projects HIP, Grant Agreement No. 221889, iQIT, Grant Agreement No. 270843, and EQuaM, Grant Agreement No. 323714. SMG acknowledges financial support from the Austrian Science Fund (FWF-P23627-N16).
\end{acknowledgments}

\appendix

\section{Proofs of the Propositions from the paper}

\subsection{Proof of Proposition \ref{propuno}}
Suppose that $\rho_{AB}$ is a classical-quantum state as given in
Eq. (\ref{classicalquantum}).
Then any local unitary operator with eigenstates $\left\{ |i_{A}\>\right\} $
will not change the state, i.e. Eq. (\ref{eq:main}) is satisfied.
Now we will show the other direction by contradiction. Suppose that
Eq. (\ref{eq:main}) holds true for some state which is not classical-quantum
and hence can be written as
\begin{equation}
\label{stateclass}
\rho_{AB}=\sum_{i,j}|i_{A}\>\<j_{A}|\otimes L_{ij}^{B},
\end{equation}
where $L_{ij}^{B}$ are local operators on part $B$ of the system. Since $\rho_{AB}$ is not
a classical-quantum state, there must exist at least one pair of indexes $k$ and $l$
with $k\neq l$ such that $L_{kl}^{B}\neq 0$.
Let the corresponding local unitary operator be $U_{A}=\sum_{i}\lambda_{i}|i_{A}\>\<i_{A}|$
with eigenstates $|i_{A}\>$ and eigenvalues $\lambda_{i}$.

If Eq. (\ref{eq:main}) must be satisfied by the state in Eq. (\ref{stateclass}), then the following equation must also be satisfied:
\begin{equation}
\<k_{A}|U_{A}\rho_{AB}U_{A}^{\dagger}|l_{A}\>=\<k_{A}|\rho_{AB}|l_{A}\> \, .
\label{eq:contradiction}
\end{equation}
The left hand side of this equation becomes $\<k_{A}|U_{A}\rho_{AB}U_{A}^{\dagger}|l_{A}\>=\lambda_{k}\lambda_{l}^{\star}L_{kl}^{B}$,
while the right hand side can be written as $\<k_{A}|\rho_{AB}|l_{A}\>=L_{kl}^{B}$.
On the other hand, since $U_{A}$ has a nondegenerate spectrum, it follows that Eq. (\ref{eq:contradiction}) can never be satisfied.

\subsection{ Proof of Proposition \ref{prop_belldiag}}

Recalling the formula for the discord of response $D_R$ for two-qubit states
given by Eq. (\ref{simplification}), we need to compute the eigenvalues of $\rho_{\gamma}U_A$
using the notation of  Eq. (\ref{matrix}) and the decomposition
\begin{equation}
U_A=\sin{\phi}\cos{\theta}\sigma_x+\sin{\phi}\cos{\theta}\sigma_y+\cos{\theta}\sigma_z.
\label{unitaryparam}
\end{equation}
The calculation of the eigenvalues of $\rho_{\gamma}U_A\equiv\rho_{\gamma}(U_A\otimes  I_B)$ is
straightforward:
\begin{eqnarray}
M&\equiv&\sum_i\left|\xi_i\left(\rho_{\gamma}U_A\right)\right|\\\
&=&\sqrt{2} \left(\sqrt{\left|Z(\phi,\theta)-\sqrt{Z(\phi,\theta)^2-4 \gamma_1  \gamma_2  \gamma_3  \gamma_4
   }\right|}\right.\nonumber\\
&+&\left.\sqrt{\left|Z(\phi,\theta)+\sqrt{Z(\phi,\theta)^2-4 \gamma_1  \gamma_2  \gamma_3  \gamma_4
   }\right|}\right),
\label{formula}
\end{eqnarray}
where
\begin{eqnarray}
Z(\phi,\theta)&=&(\gamma_1  \gamma_3 +\gamma_2  \gamma_4 ) \cos ^2(\phi )\cos ^2(\theta )\\
&&\!\!\!\!\!\!\!\!\!\!\!\!\!\!\!\!\!\!+(\gamma_2  \gamma_3
   +\gamma_1  \gamma_4 ) \sin ^2(\phi ) \cos ^2(\theta )+(\gamma_1
   \gamma_2 +\gamma_3  \gamma_4 ) \sin ^2(\theta ).
   \nonumber
\end{eqnarray}
Next, we notice that we can dismiss the absolute values due to the positivity of the inner expressions,
which is guaranteed by the following inequality:
\begin{equation}
Z(\phi,\theta)^2\geq4 \gamma_1  \gamma_2  \gamma_3  \gamma_4 \, .
\end{equation}
\begin{proof}
Proof of the above inequality goes as follows. Without loss of generality, we can assume that
$(\gamma_1  \gamma_3 +\gamma_2  \gamma_4 )\geq(\gamma_2  \gamma_3 +\gamma_1   \gamma_4 )$ and $(\gamma_1
   \gamma_2 +\gamma_3  \gamma_4 )\geq(\gamma_2  \gamma_3 +\gamma_1   \gamma_4 )$. Then:
\begin{eqnarray}
Z(\phi,\theta)&=&(\gamma_1  \gamma_3 +\gamma_2  \gamma_4 ) \cos ^2(\phi )\cos ^2(\theta )\\
&&\!\!\!\!\!\!\!\!\!\!\!\!\!\!\!\!\!\!+(\gamma_2  \gamma_3
   +\gamma_1  \gamma_4 ) \sin ^2(\phi ) \cos ^2(\theta )+(\gamma_1
   \gamma_2 +\gamma_3  \gamma_4 ) \sin ^2(\theta )\nonumber\\
&\geq& (\gamma_2  \gamma_3 +\gamma_1   \gamma_4 )\geq 2\sqrt{\gamma_1  \gamma_2  \gamma_3  \gamma_4},
\nonumber
\end{eqnarray}
and we can omit the absolute values in Eq. (\ref{formula}).
\end{proof}
Finally, to determine the maximum of $M$ as a function of $\phi$ and $\theta$, we evaluate the first derivatives:
\begin{eqnarray}
\frac{dM}{d\phi}&=&\frac{dM}{dZ}\frac{dZ}{d\phi}=\frac{1}{N}(\gamma_2 -\gamma_1 ) (\gamma_3 -\gamma_4 ) \sin (2 \phi )\cos^2{\theta}\nonumber\\
\frac{dM}{d\theta}&=&\frac{dM}{dZ}\frac{dZ}{d\theta}
=\frac{1}{N}\sin {(2 \theta)} \left[\gamma_1  \gamma_2 +\gamma_3  \gamma_4\right.\\
&-&\left.(\gamma_1  \gamma_3 +\gamma_2  \gamma_4 ) \cos ^2(\phi )-(\gamma_2  \gamma_3 +\gamma_1
    \gamma_4 ) \sin ^2(\phi )\right].\nonumber
\end{eqnarray}
where
$$N=\sqrt{Z-\sqrt{Z^2-4 \gamma_1  \gamma_2  \gamma_3  \gamma_4}}+\sqrt{Z+\sqrt{Z^2-4 \gamma_1  \gamma_2  \gamma_3  \gamma_4 }}.$$
We observe that the denominators cannot vanish, so that the extremes are determined only if the numerators vanish, i.e. for $\theta=\frac{\pi}{2}$, or $\{\theta=0,\phi=0\}$, or $\{\theta=0,\phi=\frac{\pi}{2}\}$.
Whether an extremum is a maximum or a minimum depends on the ratios of eigenvalues of $\rho_{\gamma}$.
Substituting the corresponding values of $\phi$ in $M$ and taking $1-M$ we recover Proposition \ref{prop_belldiag}.

\subsection{ Proof of Proposition \ref{proptr2}}\label{appprop2}

Denote by $S=sp(U_A)$ the arbitrary nondegenerate spectrum of one-qubit unitary transformations $U_A$,
while by $S_h$ denote the harmonic spectrum $\{1,-1\}$.
We show that for arbitrary two-qubit state $\rho_{AB}$
\begin{eqnarray}
\forall_{S},&&\  \min_{U_{A}:\,sp(U_S)=S}\|\rho_{AB}-U_{A}\rho_{AB}U_{A}^{\dagger}\|_{Tr}\label{proof1ineq1}\\
&&\ \ \ \ \ \ \ \ \ \
\leq\min_{U_A:\,sp(U_A)=S_h}\|\rho_{AB}-U_{A}\rho_{AB}U_{A}^{\dagger}\|_{Tr} \, . \nonumber
\end{eqnarray}
By $D_{A}^{(S)}$ and $D^{(h)}_A$ we denote, respectively, the diagonal matrices with spectrum $S$ and with harmonic spectrum $S_h$.
We use the property that $\min_{U_A:\,sp(U_A)=S}\|\rho_{AB}-U_A\rho_{AB}U_A^{\dagger}\|_{Tr}
= \min_{V_A}\|V_A\rho_{AB}V_A^{\dagger}-D^{(S)}_AV\rho_{AB}V_A^{\dagger}D^{(S)\dagger}_A\|_{Tr}$, where $V_A$ can be any one-qubit unitary transformation.
Assume that $\rho_{AB}'$ is a state $V_A\rho_{AB}V_A^{\dagger}$ which minimizes the above formula.
We show that for any spectrum different than the harmonic spectrum there exists the distance $\|\rho_{AB}'-D^{(S)}_A\rho_{AB}'D_{A}^{(S)\dagger}\|_{Tr}$ which is smaller or equal than  $\|\rho_{AB}'-D^{(h)}_A\rho_{AB}'D^{(h)\dagger}_A\|_{Tr}$.

Without lost of generality we assume that $S=\{e^{i\omega},e^{-i\omega}\}$.
Let us represent state $\rho_{AB}'$ as $\sum_{i,j=1}^2\ket{i_A}\bra{j_A}\otimes L^{ij}_B$, where $L^{ij}_B$ are matrices of dimensions $d_B\times d_B$. Here $\{\ket{i_A}\}_{i=1}^2$ is a basis in which matrix $D^{(S)}_A$ is diagonal.
 Let us estimate the following distance
\begin{eqnarray}
&&\!\!\!\!\!\!\!\!\!\!\!\!\|\rho_{AB}'-D^{(S)}_A\rho_{AB}'D^{(S)\dagger}_A\|_{Tr}=
\|\rho_{AB}'D^{(S)}_A-D^{(S)}_A\rho_{AB}'\|_{Tr}\\&=&
\left\|\begin{pmatrix}0&L^{12}_B(e^{i\omega}-e^{-i\omega})\\ L^{21}_B(-e^{i\omega}+e^{-i\omega})&0\end{pmatrix}\right\|_{Tr}\\&=&
2\left\|\begin{pmatrix}0&L^{12}_B\sin{\omega}\\ -L^{21}_B\sin{\omega}&0\end{pmatrix}\right\|_{Tr}\\&=&
2|\sin{\omega}|\left\|\begin{pmatrix}0&L^{12}_B\\ L^{21}_B&0\end{pmatrix}\right\|_{Tr}\leq
2\left\|\begin{pmatrix}0&L^{12}_B\\ L^{21}_B&0\end{pmatrix}\right\|_{Tr}.
\end{eqnarray}
The inequality is saturated for $\omega=\frac{\pi}{2}$, that is the case of the harmonic spectrum of $D_A^{(S)}$.
This fact proves Ineq. (\ref{proof1ineq1}) and completes the proof of Proposition \ref{proptr2}.

\end{document}